\documentclass[aps,prl,reprint,groupedaddress]{revtex4-1}

\usepackage{graphicx}
\usepackage{dcolumn}
\usepackage{bm}
\usepackage{amsmath}
\usepackage{epsfig}
\DeclareMathOperator{\Tr}{Tr}
\DeclareMathOperator{\sech}{sech}

\begin{document}


\title{Low-Weight Pauli Hamiltonian Sequences for Noise-Resilient Quantum Gates}


\author{Ryan J. Epstein}
\email[]{ryan.epstein@ngc.com}
\affiliation{Northrop Grumman Corporation, Aurora, Colorado, 80017, USA}


\date{\today}

\begin{abstract}
A simple protocol based on low-weight Pauli Hamiltonians is introduced for performing quantum gates that are robust to control noise.  Gates are implemented by an adiabatic sequence of single-qubit fields and two-qubit interactions with a single ancillary qubit, whereas related techniques require three-qubit interactions, perturbation gadgets, higher dimensional subsystems, and/or more ancilla qubits.  Low-weight interactions and low qubit overhead open a viable path to experimental investigation, while operation in a degenerate ground space allows for physical qubit designs that are immune to energy relaxation.  Simulations indicate that two-qubit gate error due to control noise can be as low as $10^{-5}$, for realizable coupling strengths and time-scales, with low-frequency noise that is as high as 15\% of the control pulse amplitude.
\end{abstract}


\maketitle


\section{introduction}

Building quantum computing hardware with increasing degrees of complexity is a considerable challenge.  While there has been steady progress in scaling to larger numbers of qubits with superconducting circuits \cite{barends,weight4}, for example, developing components with higher tolerance to noise could accelerate further scaling.  In many implementations, energy splittings between qubit states make coherence times sensitive to energy relaxation processes, such as dielectric loss \cite{loss1, loss2}, in addition to causing continuous precession of each qubit's state vector which must be accounted for \cite{calibration}.  To perform gates, resonant fields or interactions are typically applied for a period of time and quantum phase accumulates \textit{linearly} with control pulse duration and/or amplitude. This makes gate error particularly sensitive to deviations from the ideal control signals.  The goal of controlling many qubits in a large-scale system motivates the search for alternative gate implementations.  

Here, we introduce a protocol based on sequences of low-weight Pauli Hamiltonians \cite{patentDD} that is remarkably tolerant to timing and amplitude noise.  These gates are determined by the \textit{sequence} of control pulses -- that turn on single-qubit fields and two-qubit interactions -- not their duration or strength.  In addition, energy levels employed here are nominally degenerate, so that undesired phase accumulation is minimized, and there is no need for resonant driving of transitions using electromagnetic fields.  This scheme simplifies Adiabatic Gate Teleportation (AGT), a protocol introduced by Bacon and Flammia \cite{agt}. Computation is performed in a degenerate ground space with an energy gap to excited states, differentiating it from some holonomic methods \cite{zhang}, and eliminating energy relaxation as a fundamental limit to qubit coherence time. Adiabatic interpolation between Hamiltonians provides the robustness to control pulse variations.  Whereas AGT interpolates between two Hamiltonians per gate, here we employ a sequence of 3 or more.  Use of longer sequences enables a no-go argument \cite{agt} precluding reduction of ancilla qubit overhead to be circumvented. Thus, the number of ancilla qubits for two-qubit gates is reduced from 4 to 1, and the ``weight'' of qubit interactions lowered from 3-qubit to 2-qubit.  Other approaches to performing adiabatic gates in a degenerate ground space rely on 3-qubit interactions \cite{hen, oreshkov}, three-level subsystems \cite{sjoqvist, renes}, careful balancing of interaction strengths \cite{chancellor}, or perturbation gadgets \cite{kempe}. Like holonomic gate methods \cite{holonomic, oreshkov}, our gate protocol exploits a higher dimensional space than the qubit subspace in order to manipulate quantum information and maintain degeneracy. At least one ancilla qubit is needed to enlarge state space without resorting to higher dimensional subsystems.  While additional states provide an avenue for probability amplitude to leak out of the qubit subspace, such leakage error is mitigated by use of a large energy gap and is detectable by measuring the ancilla qubit's state.  This approach is somewhat more general than holonomic gates on qubits \cite{oreshkov} in that cyclic adiabatic evolutions are not necessary \cite{kult}. Ancilla qubits are only entangled with data qubits in the middle of the gate and therefore need not start and end in the same states.  

There is a noteworthy connection between gates based on adiabatic Hamiltonian interpolation and particular measurement-based gate schemes \cite{leung, leungArx, childs}. An adiabatic gate can be converted to a measurement-based variant by treating the Hamiltonian sequences, such as those described below, as measurement operators.  The two approaches differ in that tunable two-qubit interactions are directly implemented in, e.g., superconducting hardware \cite{harris}, whereas two-qubit measurements are typically constructed from single- and two-qubit gates, single-qubit measurements and additional ancilla qubits.  Moreover, adiabatic evolution avoids the random Pauli operator corrections inherent in measurement-based gates \cite{childs}. 

\section{\label{protocol} adiabatic gate protocol}

Here, we describe a universal set of gates and present simulations of a \textsc{cnot} gate with noisy control pulses. The Hamiltonians are of the form $H_{\text{tot}}= \sum_{i}H_i$, where $H_i=-g_i(t)P_i/2$, $P_i$ is a tensor product of Pauli operators, $t$ is time, and the strengths $g_i(t)\geq0$ are determined by external control signals. We use the shorthand  $H_1 \rightarrow H_2 \rightarrow H_3$ to represent a two-leg adiabatic sequence. The negative sign in $H_i$ sets the ground states to be the positive eigenvalues of $P_i$. The sequence of $P_i$ and time-dependence of $g_i$ ensure that $H_{\text{tot}}$ has a degenerate ground space and an energy gap to unused states at all times. The temporal profiles of $g_i$ are such that each turn on and off sequentially, there is a period of time wherein each $g_i$ is non-zero while the others are zero, the respective down and up ramps of $g_i$ and $g_{i+1}$ overlap in time, and the ramps are slow relative to the gap.  With only two $g_i$ being non-zero at a time, the gap is given by $(g_i^2+g_{i+1}^2)^{1/2}$. Thus, if initially prepared in the ground space, the system will remain there with high probability due to quasi-adiabatic evolution. 

Consider the case where $P_1=X_2$, $P_2=Z_1Z_2$, and $P_3 = X_1$. Here $X$ and $Z$ are Pauli operators and the tensor product in $Z_1Z_2$ is implicit.  $H_1$ has two ground states, $|0+\rangle$ and $|1+\rangle$, with $|+\rangle=(|0\rangle+|1\rangle)/\sqrt{2}$.  Clearly, when ``encoding'' a qubit in those two states, the quantum information resides exclusively in qubit 1.  Similarly for $H_3$, the quantum information resides solely in qubit 2. For $H_2$, however, the two ground states are $|00\rangle$ and $|11\rangle$, so the encoded information is spread across both physical qubits. 

\begin{figure}
	
	\includegraphics{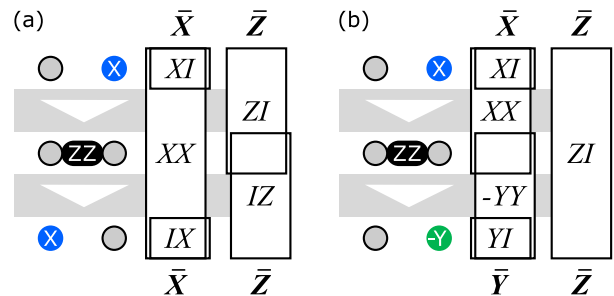}
	\caption{\label{fig:z_rot} Summary of (a) $\textsc{move}$ and (b) $S$ gates.  Time flows from top to bottom.  Hamiltonian terms are shown pictorially on the left.  Qubits are depicted as circles.  In the gray bands, the Hamiltonian is a linear combination of terms above and below the band. Logical operators are tracked on the right. Boxes encompass regions where an associated operator commutes with the total Hamiltonian. Bold logical operator labels at top and bottom identify operator behavior at the beginning and end of the gate respectively.}
\end{figure}

One might already guess that this sequence performs a ``\textsc{move}'' gate, sending the quantum information from qubit 1 to qubit 2.  We show this more rigorously by defining logical operators that act on the encoded data and tracking their evolution, as summarized in Fig.~\ref{fig:z_rot}(a).  The transformation of logical X and Z operators uniquely defines the gate operation.  For the example sequence, it is natural to choose $\bar{X}=X_1$ and $\bar{Z}=Z_1$ as logical operators when only $H_1$ is non-zero. We then construct an equivalent operator by multiplying $\bar{X}$ by $P_1$ ($=X_2$) to give $\bar{X}'=X_1X_2$.  This new operator still commutes with $H_1$ and acts the same way as $\bar{X}$ on the ground states. However, $\bar{X}'$ and $\bar{Z}$ commute with $H_{\text{tot}}$ for the entire leg $H_1 \rightarrow H_2$, whereas $\bar{X}$ does not commute with $H_2$.  Next, when only $H_2$ is non-zero, we multiply $\bar{Z}$ by $P_2$ to obtain $\bar{Z}'=Z_2$. Both logical $Z$ operators act equivalently on the ground space when just $H_2$ is non-zero, but now $\bar{Z}'$ commutes with $H_{\text{tot}}$ for the second leg, $H_2 \rightarrow H_3$.  These operator ``handoffs'', $\bar{X}$ to $\bar{X}'$ and $\bar{Z}$ to $\bar{Z}'$, are an important feature of our gate scheme.  Finally, when just $H_3$ is non-zero, we multiply $\bar{X}'$ by $P_3$, producing $\bar{X}''=X_2$.  We see that the logical operators are transformed as $X_1\rightarrow X_2$ and $Z_1\rightarrow Z_2$.  This \textsc{move} gate is closely related to a measurement-based implementation of one-bit teleportation \cite{onebittele, childs}.

Next, consider what happens when $P_3$ is replaced with $-Y_2$, as in Fig.~\ref{fig:z_rot}(b).  The first leg, $H_1\rightarrow H_2$, is the same as before.  When only $H_2$ is non-zero, however, a new equivalent logical $X$ operator is $-Y_1Y_2$ (obtained by multiplying $X_1X_2$ with $P_2$), which commutes with the second leg $H_2\rightarrow H_3$. When only $H_3$ is non-zero, we multiply $-Y_1Y_2$ by $P_3$ to give $Y_1$.  At the end of the gate, the encoded information is localized in qubit 1 and qubit 2 is in the state $(|0\rangle-i|1\rangle)/\sqrt{2}$. This sequence has caused the logical operators to transform as $Z_1\rightarrow Z_1$ and $X_1\rightarrow Y_1$, which is a 90 degree rotation around the $\hat{z}$ axis (the $S$ gate). Note that by choosing to apply the $-Y$ field on qubit 2, the quantum information started and ended on qubit 1.  No ``teleportation'' was necessary for the gate to occur. 

Rotations about $\hat{z}$ by an arbitrary angle $\theta$ are implemented by replacing $H_3$ with $A_2^T=cX-sY$, where $c=\cos(\theta)$ and $s=\sin(\theta)$. To see this, let us construct the logical operator $\bar{A}^T=c\bar{X}-s\bar{Y}=cX_1X_2-sX_1Y_2=X_1A_2^T$. This operator commutes with $H_2$ and $H_3$. When just $H_3$ is non-zero, the operator acts like a logical X since the quantum information resides only in qubit 1, and there are only X terms acting on qubit 1 in $c\bar{X}-s\bar{Y}$.  Therefore, we have the operator transformation $c\bar{X}-s\bar{Y}\rightarrow \bar{X}$.  Since $Z$ is not transformed, we identify this gate as the desired $\hat{z}$ rotation by $\theta$.  It should also be clear that suitably permuting Pauli operators allows for $\hat{x}$ or $\hat{y}$ rotations.

While the above protocols enable arbitrary single qubit gates, it is worth noting a Hadamard gate variation using the sequence $X_2 \rightarrow X_1Z_2 \rightarrow Z_1$. The same method for determining operator transformations is applicable here. For this gate, the quantum information necessarily starts in one qubit and ends in the other.  If desired, one could move the information back using the \textsc{move} gate.
 
\begin{figure}
	\includegraphics{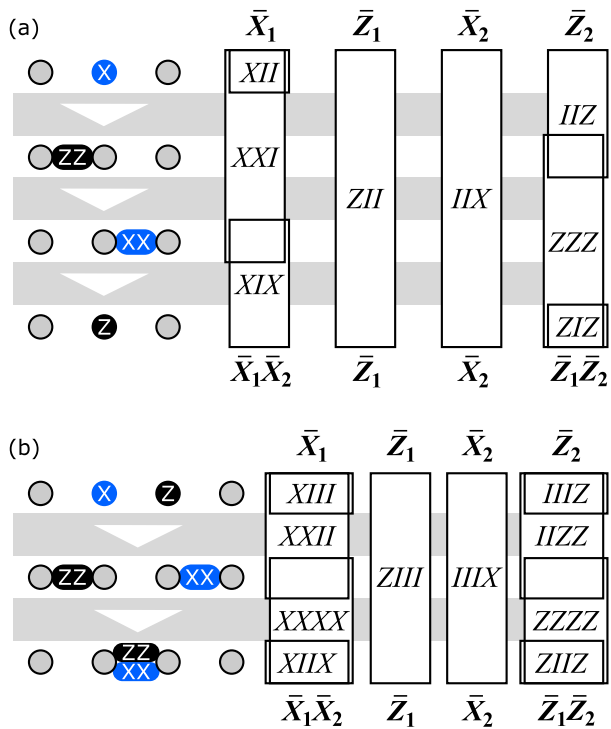}
	\caption{\label{fig:cnot} Summary of \textsc{cnot} gates using one (a) and two (b) ancilla qubits.  The diagrams are interpreted as in Fig.~\ref{fig:z_rot}.}
\end{figure}

\begin{figure*}
	\includegraphics{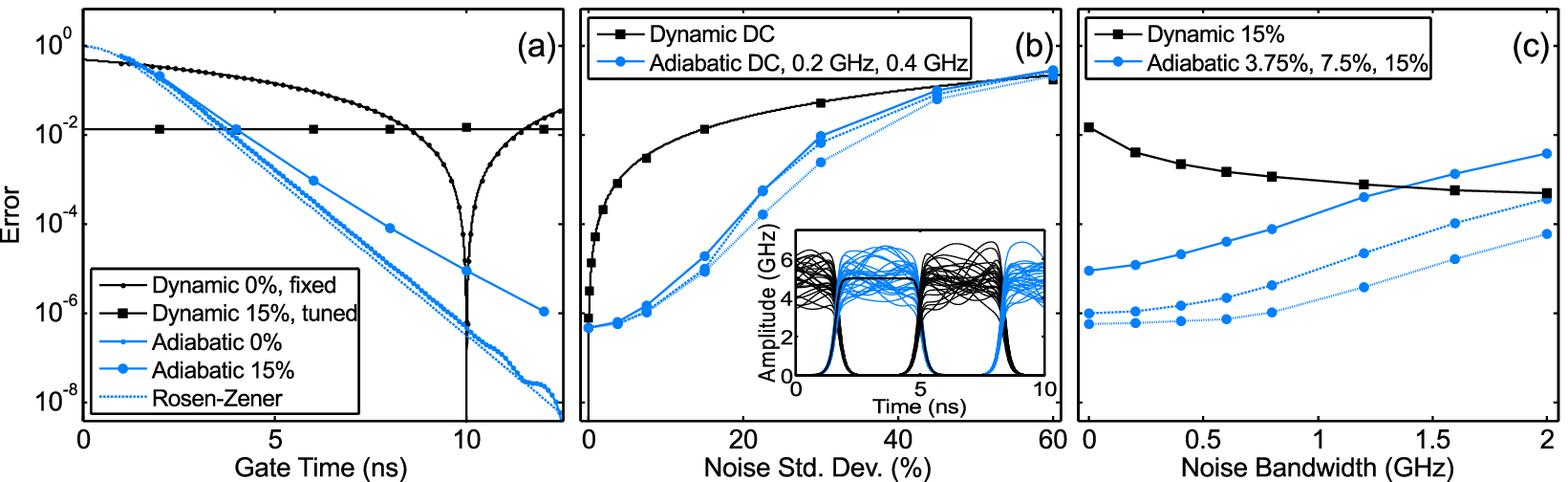}
	\caption{\label{fig:sim} Numerical simulations of the 3-qubit adiabatic \textsc{cnot} gate and a two-qubit dynamic \textsc{cnot} gate. (a) Gate error versus gate time without noise (0\%) and with 15\% DC amplitude fluctuations. The solid black curve is the analytic error for fixed pulse amplitude (labeled ``fixed''). The solid black line is the analytic error for the dynamic gate with pulse amplitude tuned for minimal error and $\sigma_f = 0.15$.  The dashed line is the Rosen-Zener transition probability, $\sech^2(\pi grt_g)$, with no free parameters, where $g=5$ is the gap, and $r=0.052$ is the ratio of the Rosen-Zener time constant to the gate time used in the gate pulses. (b)  Gate error versus noise standard deviation with 10 ns gate time for DC amplitude fluctuations (DC; dotted line) and for noise bandwidths of 0.2 GHz (dashed line) and 0.4 GHz (solid line). The inset shows several instances of pulses with 15\% noise, and one without noise. (c) Gate error versus noise bandwidth for 10 ns gate time and noise amplitudes of 3.75\% (dotted line), 7.5\% (dashed line) and 15\% (solid line). For all plots, gate error is calculated as $1-|\Tr(U_{\text{ideal}}^\dag O_{\text{sim}})|^2/d^2$, where $d=4$ is the subspace dimension, $U_{\text{ideal}}$ is the ideal \textsc{cnot} operator, and $O_{\text{sim}}$ is the simulated operator acting on the ground subspace with matrix elements $\langle j|U_{\text{sim}}|i\rangle$.  Here, $|i\rangle$ and $|j\rangle$ are the initial and final basis states, respectively, and $U_{\text{sim}}$ is the simulated unitary acting on the full state space. Noise is created from a vector of gaussian-distributed pseudo-random numbers with a 4th order low-pass filter applied in the Fourier domain. Bandwidth denotes the cutoff frequency of the filter. Each data point with noise applied is the average of 1000 Monte Carlo simulations.  Note that the simulation results remain valid if time and frequency units are changed consistently, e.g. ns $\rightarrow$ $\mu$s and GHz $\rightarrow$ MHz.}
\end{figure*}

We have also constructed two-qubit gate sequences, such as the \textsc{cnot} gate sequence depicted in Fig.~\ref{fig:cnot}(a). This gate is the most significant of the ones presented due to its marked simplification compared to AGT \cite{agt}, reducing the number of ancilla qubits from 4 to 1, and the interaction weights from 3-qubit to only 2-qubit.  This gate is the focus of the simulations discussed below.  Another \textsc{cnot} gate is shown in Fig.~\ref{fig:cnot}(b).  This \textsc{cnot} sequence has only two legs but does require two ancilla qubits instead of one.  Note that the \textsc{cnot} gates are transformed into controlled-Z gates by replacing $XX$ in the penultimate $H_i$ with $XZ$.

\section{\label{sims} \textsc{cnot} gate simulations}

We now turn to gate simulations.  Figure~\ref{fig:sim} details results from simulations comparing the three-qubit adiabatic \textsc{cnot} gate and a standard two-qubit ``dynamic'' \textsc{cnot} gate.  For the latter, a $ZZ$ coupling is turned on for a period of time to dynamically generate an entangling gate which, in the absence of noise, is equivalent to a \textsc{cnot} gate up to (omitted) single qubit rotations. A given simulation run consists of unitary evolution with or without an instance of classical noise applied to the control signals.  No other source of decoherence is included. 

Figure~\ref{fig:sim}(a) is a plot of gate error as a function of gate duration. For the noise-free case the pulse amplitudes are fixed, and we see that gate error decreases exponentially with gate time for the adiabatic gates, whereas for the dynamic gate, error is only below $10^{-4}$ in a small range around 10 ns.  An analytic expression for the dynamic gate error, $1-\cos(\pi(t_g/t_0-1)/4)$, agrees with the simulated data, where $t_g$ is gate time and the pulse amplitude has been manually set so that $t_0=10$ ns. These results exhibit an important benefit of adiabatic gates: no fine-tuning of pulse amplitudes, durations, or timing is required.

For the simulations presented here we have chosen a special pulse shape in order to compare to an analytic result of Rosen and Zener \cite{rosen}. They analyzed a Stern-Gerlach experiment in which a spin-1/2 is subjected to a magnetic field that rotates from one angle to another at an angular rate $\dot{\theta}\propto\sech(\pi t/\tau)$, where $t$ is time and $\tau$ is the Rosen-Zener time constant. When the total rotation angle is 90 degrees ($\theta: 0 \rightarrow \pi/2$), the field can be described by the interpolation $\cos(\theta)Z+\sin(\theta)X$, which is analogous to our gate interpolations.  Here, we use the same rate of change of the effective field angle and the time constant is a fixed fraction of the gate time.  The resulting pulse shape is shown in the inset of Fig.~\ref{fig:sim}(b).  The error for the Rosen-Zener case is the probability of being in the excited state. We find that the exponential decay rate of error as a function of gate time is the same for both the Rosen-Zener problem and our adiabatic gates. The two problems are related in that $Z\rightarrow X$ is similar to $P_i\rightarrow P_j$, where $P_i$ and $P_j$ are two generic anti-commuting Pauli products. The differences are that our gate Hamiltonians have a degenerate ground space in addition to a larger state space.  The small oscillations in the adiabatic gate error simulations at gate times above 10~ns are due to finite pulse length effects.  It should be emphasized that the Rosen-Zener pulse shape was not chosen so as to reduce error; it is not necessary to use a particular pulse shape to achieve low errors.

We also plot in Fig.~\ref{fig:sim}(a) gate error with independent shot-to-shot variation in the pulse amplitudes, which we call ``DC noise'' since the pulse shape does not change.  For the dynamic gate, the pulse amplitude is adjusted to minimize gate error at each gate time.  In this case, the gate error is above 1\% and independent of the gate time, consistent with the analytic result, $\pi^2\sigma_{f}^2/16$.  This expression is valid to second order in $\sigma_f$, the standard deviation of the noise as a fraction of pulse amplitude.  Remarkably, for the adiabatic gate, errors below $10^{-5}$ are achieved for 10 ns gate time and 5 GHz average energy gap (divided by Planck's constant) with fluctuations as high as 15\% of the control pulse amplitude.  This energy gap is much larger than the 40 MHz interaction strength needed for the dynamic gate, and was selected to show error scaling down to $10^{-8}$.  In a separate simulation (not shown), we achieved a gate error below $10^{-4}$ with 12\% noise and a 1.6~GHz energy gap, using pulses constructed from Gauss error functions. Increasing the gate time by a factor of, say, 4 would further reduce the required gap by the same factor.

Gate error as a function of noise amplitude is plotted in Fig.~\ref{fig:sim}(b) for a gate time of 10 ns. Error for the dynamic gate with DC noise increases quadratically with noise amplitude and is well described by an analytic expression (see figure caption).  For the adiabatic gate, the error increases much more gradually both for DC noise and for low-frequency noise.  Gate error is affected by control noise due to changes in the instantaneous energy gap, and the rate of change of the Hamiltonian.  Both of these contribute to excitation out of the ground space.

Fig.~\ref{fig:sim}(c) shows gate error versus noise bandwidth for 15\% noise amplitude and a gate time of 10 ns. For each bandwidth, noise amplitude is adjusted so that the standard deviation of the amplitude at the peak of the pulse remains constant.  In this situation, the dynamic gate error actually decreases with increasing bandwidth because the effect of high-frequency noise is averaged out over the duration of the pulse while low-frequency noise power decreases.  The lack of ancillary energy levels to excite is also a beneficial factor.  For the adiabatic gate, the error increases approximately exponentially with noise bandwidth above 0.8 GHz for the three noise amplitudes plotted.  As bandwidths decrease below 0.8 GHz, the errors for the two smaller noise amplitudes level off near the noise-free error.  Higher frequency noise components lead to fluctuations in the rate of change of the Hamiltonian, causing more leakage error. While leakage can be problematic for quantum error correction if it is undetected, a beneficial feature of this approach is that leakage errors are detectable by measuring the state(s) of the ancilla qubit(s).  If they are measured to be in their excited states at the end of the gate, a leakage error has occurred.

The inset of Fig.~\ref{fig:sim}(b) presents several instances of $g_i(t)$ pulses with 15\% noise of 0.2 GHz bandwidth. The noise is \textit{multiplied} by the pulses, making noise amplitude proportional to pulse amplitude.  This was done to highlight the way in which these gates are effective: field and interaction strengths can be very noisy when turned on but must turn off strongly.  In a physical system, it would be desirable for the qubit energy levels to split nonlinearly so as to suppress noise when the control signal is low.  This type of behavior occurs in a superconducting flux qubit with a flux-tunable tunnel barrier \cite{mooij, orlando, paauw}.  If the barrier is raised with a control flux beyond the point where tunneling (an $X$ field) is shut off, noise of sufficiently low frequency and amplitude does not cause splitting of the two persistent current ground states.  On the other hand, flux in the main qubit loop tilts the potential well linearly with control flux (producing a $Z$ field), which is undesirable. Even with physical qubits that respond linearly to control signals, however, a nonlinear response can be created in composite qubits comprised of several physical qubits coupled together \cite{patentPNS}.

Let us briefly address imperfections in the ideal Pauli term $P_i$ of the form $P_i+\epsilon P_j$, where $P_j$ is undesired and $\epsilon$ is small.  It is found in separate simulations without noise (not shown) that $P_1 = X_2+\epsilon Z_2$ and $P_3=X_2X_3+\epsilon Y_2Y_3$ both give rise to gate error that is quadratic in $\epsilon$, and that the respective values of $\epsilon$ are of order $10^{-3}$ and $10^{-4}$ to produce gate errors below $10^{-5}$.  Creating interactions with low imperfection is the key condition for the effectiveness of the proposed scheme.  In a physical implementation, it would be highly beneficial for these interactions to be defined via device geometry or topology so that robustness is built in during fabrication.  For flux qubits, tunable $X$, $Z$ and $ZZ$ terms have all been demonstrated in experiments \cite{paauw, harris}.  An $XX$ interaction with fully independent tuning has yet to be demonstrated, but designs have been developed \cite{patentXX}.  Simultaneously demonstrating all four interactions for the \textsc{cnot} gate with high purity is a significant experimental challenge.

\section{conclusions}

We have presented a quantum gate protocol that uses low-weight interactions and achieves low two-qubit gate errors with quite noisy control pulses.  In addition, control pulse duration only weakly influences gate error via leakage at short times and decoherence at long times relative to the energy gap, the latter effect not being included in the presented simulations.  On the other hand, the simple gate variations presented here require much stronger interactions than standard gates, and do not protect against local noise that can split the ground space and cause dephasing.  These methods have led to a more complete solution that counteracts local noise by encoding qubits in noise-suppressing Hamiltonians while maintaining robustness to control noise and use of two-qubit interactions for all gates \cite{patentPNS}.  We anticipate this lowering the barrier to experimental implementation when compared to approaches that require higher weight interactions \cite{oreshkovFTHQC,oreshkovPRA,brun2014,brun2015,cesare,marvian}.

\begin{acknowledgments}
The author wishes to thank D. G. Ferguson, B. Eastin, W. G. Brown, P. D. Nation, and J. T. Anderson for illuminating discussions and critical review of the manuscript, and D. Bacon for pointing out some of the measurement-based gate work.
\end{acknowledgments}

\bibliography{double_dragon}

\end{document}